\documentclass{article}

\usepackage[dblblindworkshop, final]{neurips_2025}
\usepackage{subcaption}
\workshoptitle{Generative AI in Finance}

\usepackage[utf8]{inputenc} 
\usepackage[T1]{fontenc}    
\usepackage{hyperref}       
\usepackage{url}            
\usepackage{booktabs}       
\usepackage{amsfonts}       
\usepackage{nicefrac}       
\usepackage{microtype}      
\usepackage{xcolor}         
\usepackage{graphicx}
\usepackage{algorithm}
\usepackage{algpseudocode}
\usepackage{mdframed} 

\usepackage{enumitem} 
\newlength{\algoboxrulewidth}
\title{Data-driven Feynman–Kac Discovery with Applications to Prediction and Data Generation}

\author{%
  Qi Feng \\
  Department of Mathematics\\
  Florida State University\\
  Tallahassee, FL 32306 \\
  \texttt{qfeng2@fsu.edu} \\
  \And
  Guang Lin \\
  Department of Mathematics\\
  Purdue University\\
  West Lafayette, IN 47907\\
  \texttt{guanglin@purdue.edu} \\
  \AND
  Purav Matlia \\
  Department of Computer Science\\
  Purdue University\\
  West Lafayette, IN 47907\\
  \texttt{pmatlia@purdue.edu} \\
  \And
  Denny Serdarevic \\
  Department of Mathematics\\
  Florida State University\\
  Tallahassee, FL 32306 \\
  \texttt{ds22ck@fsu.edu} \\
}

\begin{document}
\maketitle
\begin{abstract}
 In this paper, we propose a novel data-driven framework for discovering probabilistic laws underlying the Feynman–Kac formula. Specifically, we introduce the first stochastic SINDy method formulated under the risk-neutral probability measure to recover the backward stochastic differential equation (BSDE) from a single pair of stock and option trajectories. Unlike existing approaches to identifying stochastic differential equations—which typically require ergodicity—our framework leverages the risk-neutral measure, thereby eliminating the ergodicity assumption and enabling BSDE recovery from limited financial time series data. Using this algorithm, we are able not only to make forward-looking predictions but also to generate new synthetic data paths consistent with the underlying probabilistic law.
\end{abstract}

\section{Introduction}
Identifying the governing laws in physical systems is of high interest in various fields, including fluid dynamics~\cite{bai2017data}, plasma dynamics~\cite{kaptanoglu2023sparse}, nonlinear optics~\cite{ermolaev2022data}, mesoscale ocean closures~\cite{zanna2020data}, and computational chemistry~\cite{harirchi2020sparse}, etc. Many of these complex physical laws could be described by PDEs with proper boundary conditions, which are deterministic dynamical systems. For each specific scientific task, the detailed deterministic form of the PDE often remains unknown. In the era of big data, when rich experimental data is available, it gives rise to an opportunity to automatically transform these data into deterministic physical laws. In light of this idea, the data-driven discovery of hidden equations (mostly deterministic equations) has been enabled by the rapid progress in statistics and machine learning. 
As a toy example, we could select a set of candidate common functions containing derivatives with different orders, and call it the library $\Theta(u):=\{1, u, u_x, u^2, u u_x, u_x^2,\cdots, u_{xx}^2\}$.
The discovery of the system/PDE $\mathbf F$ can be identified as the following symbolic regression problem, 
\begin{equation}\label{regression_2}
    \begin{pmatrix}
         u_t(t_i,x_j)
    \end{pmatrix}_{i,j}=\begin{pmatrix}
 1 & u(t_i,x_j) & u_x(t_i,x_j) & u^2(t_i,x_j)  & u_x^2(t_i,x_j) & \cdots & u_{xx}^2(t_i,x_j)
    \end{pmatrix}_{i,j} \xi,
\end{equation}
where $\{u(t_i,x_j)\}_{i,j}$ represents the data of the PDE, and $t_i$ (resp. $x_j$) represents the time (resp. spatial) points. Data driven model discovery for PDE by using physics-informed learning has been studied in   \cite{chen2021physics}. To bridge the deterministic law and probability law discovery in this context, we start with a slight modification of the problem by adding a terminal condition $u(T,x)=g(x)$ instead of the initial condition at $t=0$, and simply consider the parabolic equation with linear coefficients, 
\vspace{-0.2cm}
\begin{equation}\label{heat equ}
    \mathbf F(\cdot):=u_t+rx u_x +\frac{1}{2}\sigma^2x^2u_{xx}-ru+\tilde f(t,x)=0,
\end{equation}
where $ \sigma, r, T$ are parameters and $\tilde f$ is a known function. Due to the famous Feynman-Kac formula \cite{kac1949distributions, feynman1955slow}, the solution $u(t,x)$ (if it exists) can be represented by the following conditional expectation,\vspace{-0.2cm}
\begin{equation}\label{conditional expectation}
    u(t,x)=\mathbb E^{\mathbb Q}\Big[e^{-r(T-t)}g(X_T)+\int_t^T e^{-r(s-t)} \tilde f(s,X_s)ds \Big| X_t=x\Big],
\end{equation}
under the probability measure $\mathbb Q$, where $X_t$ is the solution of the following forward in time SDE,
\begin{equation}\label{SDE}
    d X_t = r X_t dt + \sigma X_t d B^{\mathbb Q}_t, \quad \text{with initial condition}\quad X_0=x_0,
\end{equation}
and $B_t$ is Brownian motion under $\mathbb Q$. The dynamic $X_t$ under the market measure $\mathbb P$ is unknown. By martingale representation theorem, the solution $u(t,x):=Y_t$ can be further represented by the following backward stochastic differential equation(BSDE), with the pair $(Y_t, Z_t)$ being its solution,  \vspace{-0.2cm}
\begin{equation}
    \label{BSDE}
    Y_t=g(X_T)+\int_t^T [rY_s-\tilde f(s,X_s) ]ds-\int_t^TZ_s dB^{\mathbb Q}_s, \quad \text{with}\quad Y_T=g(X_T).
\end{equation}
Feynman-Kac formula is fundamental in probability theory, connecting SDE and its corresponding PDE. Feynman-Kac has been generalized to various fields including financial mathematics \cite{bjork2009arbitrage, joshi2003concepts, karatzas2014brownian, wang2022path}, mathematical physics \cite{anapolitanos2014multipolarons, lowen1988spectral, hinrichs2024feynman}, Quantum mechanics \cite{simon1979functional, caffarel1988development, hall2013quantum}, etc. More importantly, Feynman-Kac is paired with the underlying SDE model. The inverse problem has not been studied yet, which is to rediscover the Feynman-Kac formula by linking two different type of data sets collecting from $u(t,x)$ and $X_t$. In a real world application, the data $\{X_t\}_{0\le t\le T}$ could be collected from the daily stock price, and the solution $u(t,x)$ actually represents the European call option price at time $t$ with terminal time payoff $g(X_T)=\max\{ S_T-K,0\}$, strike price $K$ and maturity T. 
Our goal is to discover the relation (i.e. BSDE) from a pair of single trajectories (stock, option). 
 The significant difference compared to the discovery of the deterministic law is that we are learning PDE with a terminal condition instead of an initial condition. Compared to data-driven sparse identification of deterministic nonlinear dynamical systems (SINDy) \cite{brunton2016discovering}, and stochastic dynamical systems \cite{wanner2024higher}, our probability law discovery is based on a pair of data.  In particular, to identify the analytical form of the generator and diffusion part of the BSDE, we generalize the SINDy algorithm to stochastic SINDy under the risk-neutral probability measure. Once the BSDE is identified, combining with the risk-neutral dynamic of the stock price, we can generate new sample paths for (stock, option) pairs. This paper is organized as below. In Section 2, we present the main algorithm. In Section 3, we present the numerical algorithms for both the Black-Scholes model and real-world financial data.

\section{Algorithm}

The objective is to discover the differential form  of the BSDE in the following general form,
\begin{equation}
      Y_t=g(X_T)+\int_t^T f(s,X_s,Y_s,Z_s)ds-\int_t^TZ_s dB^{\mathbb Q}_s, \quad \text{with}\quad Y_T=g(X_T).
\end{equation}
Since the available data are discrete, the discrete-time version is considered instead:
\begin{equation}\label{diffBSDE}
    \Delta Y_{t_i} = -f(t, X_{t_i}, Y_{t_i}, Z_{t_i})\Delta t_i + Z_{t_i}\Delta B^{\mathbb Q}_{t_i},
\end{equation}
where $\Delta t_i$ is the discrete timestep, $\Delta Y_{t_i} = Y_{t_{i+1}} - Y_{t_i}$, and $\Delta B^{\mathbb Q}_{t_i}:=B^{\mathbb Q}_{t_{i+1}}-B^{\mathbb Q}_{t_{i}}$. The dataset consists of a single pair of discrete trajectories $D_u=\{(X_{t_i}, Y_{t_i})\}_{i=0}^{i=N}$ with terminal value $X_{t_N} = X_T$; $Y_{t_N}=Y_T$. To identify the dynamics, we formulate a sparse regression problem which minimizes:
\begin{equation}\label{sparse_regression}
    \|\Delta Y_{t_i}-\Theta^f \xi^f \Delta t_i  - \Theta^Z \xi^Z \Delta B_{t_i}^{\mathbb Q}\|^2.
\end{equation}
The library matrix $\Theta = \Theta(u, X_{t_i})$ consists of two partitions: $\Theta^f$, containing candidate functions for the generator $f$; and $\Theta^Z$, containing candidate functions $Z$. Both partitions have sparse vectors of coefficients $\xi^f$ and $\xi^Z$, respectively. The discovered BSDE is then used for online prediction of $Y_t$ and generation of new trajectories consistent with the dynamics.

\textbf{Step 1 (DNN training):} Numerical methods for obtaining partial derivatives are often inapplicable or erroneous when \textit{only a single pair of trajectories} is available. To address this, a Deep Neural Network (DNN) $\mathcal{N}$ is used to approximate the solution $u(t, x)$ with inputs $(t_i, X_{t_i})$ and parameters $\beta$ \cite{chen2021physics}. Accurate derivatives are obtained by introducing a physics loss and applying automatic differentiation. 

Optimization is performed using LBFGS, which updates both $\beta$ and an auxiliary parameter $\zeta$ in the physics loss. The physics loss, computed on collocation points $\mathcal{D}_c$ randomly sampled from the bounds of the observed data, measures the residual $u_t - \Psi\zeta$ where the library matrix $\Psi=\Psi(X_t, u, u_x, u_{xx})$ is reconstructed with updated derivatives after each optimization step. The total loss function is
\begin{equation}\label{loss_func}
    \mathcal{L}(\beta, \zeta; \mathcal{D}_u, \mathcal{D}_c) = \alpha \mathcal{L}_d(\beta; \mathcal{D}_u) + \gamma \mathcal{L}_p(\beta, \zeta; \mathcal{D}_c) + \delta\lVert\zeta\rVert_1
\end{equation}
where $\mathcal{L}_d$ is the data loss, $\mathcal{L}_p$ is the physics loss, and $\lVert\zeta\rVert_1$ is an L1 regularization term. Hyperparameters $\alpha, \gamma, \delta$ control the relative weighting of each term. The auxiliary parameter $\zeta$ is discarded after training; it serves solely to improve the derivative accuracy by enforcing consistency of $\mathcal{N}$ with the underlying PDE (e.g.: equation \ref{heat equ}) constraints. 

\textbf{Step 2.0 (Estimating diffusion coefficients):} Construction of the library $\Theta(u, X_{t_i})$ requires the discrete Brownian increments $\Delta B^{\mathbb Q}_t$. These increments are obtained by first estimating the diffusion coefficient $\sigma(x)$. A preliminary estimate is computed as $\sigma_{\text{noisy}} = \sqrt{{\Delta\langle X_{t_i} \rangle}/{\Delta t_i}}$ which is often noisy due to the discrete sampling of the trajectory. We denote $\Delta\langle X_{t_i} \rangle$ as the increment of the quadratic variation of the stock price $X_t$. To obtain a smoother and more accurate approximation, the noisy estimate is modeled using SINDy by solving $\sigma_{\text{noisy}}(X_{t_i})=\Phi(X_{t_i})\nu$ where $\Phi(X_{t_i})$ is a library of candidate functions and $\nu$ is the sparse solution vector of coefficients. The resulting diffusion function is discovered by applying SINDy to $\| \hat{\sigma}(X_{t_i})-\Phi(X_{t_i})\nu\|^2$ similar to \eqref{sparse_regression}. 

\textbf{Step 2.1 (Extracting $\mathbb{Q}$ Brownian motion)}: Once the smoothed diffusion function $\hat{\sigma}(x)$ is obtained, the discrete Brownian increments $\Delta B^{\mathbb Q}_{t_i}$ are calculated from the observed trajectory by rearranging the discrete form of the SDE (\ref{SDE}):
$\displaystyle  
    \Delta B^{\mathbb Q}_{t_i} = (\Delta X_{t_i} - r X_{t_i} \Delta t)/\hat{\sigma}(X_{t_i})$. The drift coefficient $r$ is 
assumed to be a fixed, and corresponds to the risk-neutral probability measure $\mathbb Q$. These extracted increments are subsequently used to construct the library matrix $\Theta$.

\textbf{Step 3 (Stochastic SINDy for BSDE):} With the library matrix $\Theta$ constructed, the discrete BSDE is identified using stochastic SINDy under probability measure $\mathbb Q$. The sparse regression problem in \eqref{sparse_regression} is solved using sparsity-promoting algorithms with regularization, in particular, Sparse Relaxed Regularized Regression (SR3) \cite{zhengSR3} is employed. The result is a sparse set of coefficients that identify the driver $f = \Theta^f\xi^f$ and $Z = \Theta^Z\xi^Z$. In the current version, we do not provide terminal boundary discovery and we hope to add it in the future.

\textbf{Step 4 (Online Prediction):} Online predictions for $Y_{t_{i+1}}$ are performed using a memory window $(X_{t_s}, Y_{t_s})_{s \in [i-\tau, i]}$ of fixed size $\tau$ spanning from $t_{i-\tau}$ to $t_{i}$.  The dynamic of $Y_t$ is backward in time, which is solved backward in time.
Different from the deep neural network algorithms \cite{han2018solving, sirignano2018dgm} with BSDE model parameter given (i.e. $f,\sigma$ fixed), we are not solving the BSDE as the model is unknown for the future time interval $[t,T]$, but rather take the random form with unknown increment of the Brownian motion $\Delta B^{\mathbb Q}_{t_i}$. With the real data, $Y_t$ is known from the market, we simply want to make a prediction for tomorrow's option price, we use the formula \eqref{diffBSDE} with $Z_t$ independent of  $\Delta B^{\mathbb Q}_{t_i}$. This prediction loop is interleaved with the neural network $\mathcal{N}(t, x; \beta)$ retraining to regenerate partial derivatives and with rediscovery of the dynamics, ensuring consistency with the most recent memory window. We make
mean prediction (mimicking $\mathbb E_t(Y_{t_{i+1}})$) and confidence intervals based on the desired confidence level by generating independent increments $\Delta B^{\mathbb Q}_{t}$. After each prediction, the memory window is shifted forward by one time step $(X_{t_s}, Y_{t_s})_{s \in [i-\tau + 1, i + 1]}$ and the dynamic rediscovery and prediction (i.e. \textbf{Step 1-4}) is repeated. For data points immediately preceding time skips caused by market opening and closing, the true value of the stock or option after the skip is used instead of a prediction. Our algorithm does not account for market opening and closing times (i.e., non-trading intervals), and we hope to add this feature in the future. 

\textbf{Step 5 (Generation):} New trajectories are generated by running multiple simulations of the stock price using {learned} forward dynamic \eqref{SDE} and using the discovered BSDE \eqref{diffBSDE} by sampling Brownian increments to create trajectories consistent with the learned dynamics with shifted window $(X_{t_s}, Y_{t_s})_{s \in [i-\tau, i]}$ as described in \textbf{Step 4}. For online prediction and data generation, our generated option paths are not option prices by solving a given BSDE model \cite{han2018solving, sirignano2018dgm} backward in time as the models (both stock and option dynamics) are unknown. We generate the option paths by iteratively updating our models and making future predictions.
\section{Examples}
\noindent \textbf{Example 1: Black-Scholes discovery.} We use the Black-Sholes model as a benchmark to test our algorithm. 
For this example, we can compare the analytical solution and its derivatives to our auto-differentiated partial derivatives from our Neural Network in \textbf{Step 1}. 
Following our algorithm, once the BSDE is trained in \textbf{Step 3}, given the current market option price $Y_t$ (assumed to be computed from an analytical solution), we can compute the next option price $Y_{t+1}$ given we know the increment of the Brownian motion. We then use this idea to do prediction and generating new data as in \textbf{Step 4-5}. To make our prediction and generation consistent with our model and analytical solution, we need to retrain the neural network with the shifting window to update more accurate values of the derivatives of $u(t,x)$. Our numerical results are consistent with our desired algorithm performance.
To generate the data pair of stock and option, we simulate the underlying asset price following the geometric Brownian motion,
$dS_t = \mu S_tdt + \sigma S_t dB^{\mathbb{P}}_t$, with $\mu=0.3$, $\sigma=0.2$.  Applying Girsanov theorem with a select risk-free interest rate $r=0.1$, we get \eqref{SDE} under $\mathbb Q$. Then, the corresponding European call options $Y_t$ are computed using the closed-form formula or the Black-Sholes PDE (i.e.  \eqref{heat equ} with $\tilde f=0$) with maturity time $T=1$ and payoff $g(S_T) = \max\{S_T - K,0\}$.  We train the model on $[0, 0.8]$ and test the model for the interval $[0.8,1]$, with a total number of time steps $N=5*10^5$. By training from \textbf{Step1-3}, our stochastic SINDy discovers the BSDE $dY_t = -[1.018(u_{t})(dt) + 0.106(S_t)(u_s)(dt) + 0.016(S^2_t)(u_{ss})(dt)] + 0.2(S_t)(u_s)(dB^{\mathbb{Q}}_t)$, where we use $(u_{t}), (u_{s}), (u_{ss})$ to represent the derivatives computed from the neural network. The numerical results for the Black-Scholes model are presented in Figure [1-(a)(b)(c)(d)].

\noindent \textbf{Example 2: Market data  Apple.} Data was collected from the ticker AAPL from 11/14/2024 up to expiration 07/18/2025 with timer intervals of 1 second at a strike price at $K=170$. We set $T=1$ with $N=3907974$ and we use an 80\%-20\% train-test split. We present the prediction in Figure [1-(e)] and generate stock-option pairs in Figure [1-(f)(g)].


\begin{figure}[htbp]
\centering
\begin{minipage}{0.55\textwidth}
\includegraphics[width=0.8\textwidth]{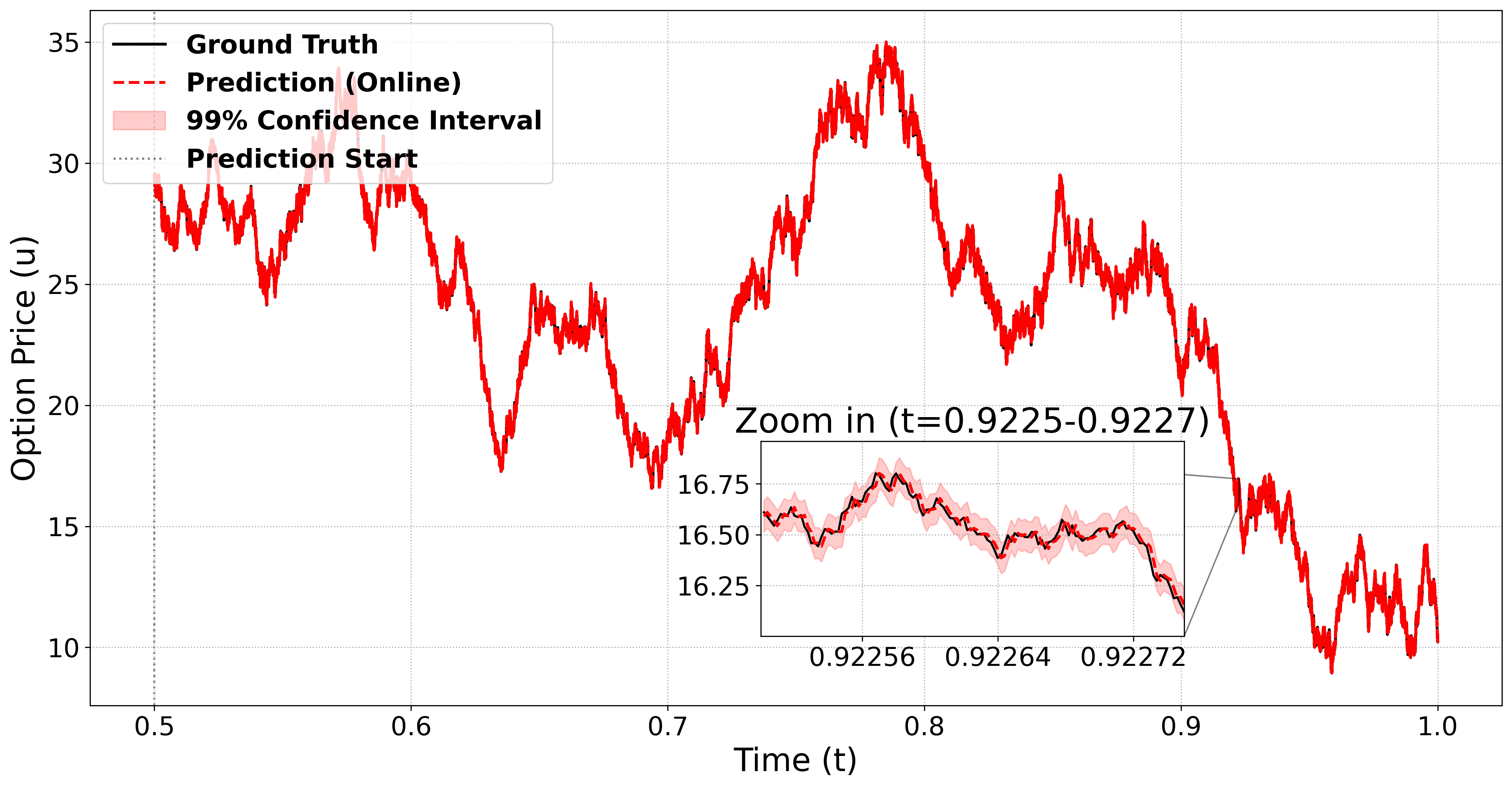}
\subcaption[second caption.]{\small  Prediction: option prices vs ground truth}\label{fig:1b}
\end{minipage}%
\begin{minipage}{0.55\textwidth}
\includegraphics[width=0.8\textwidth]{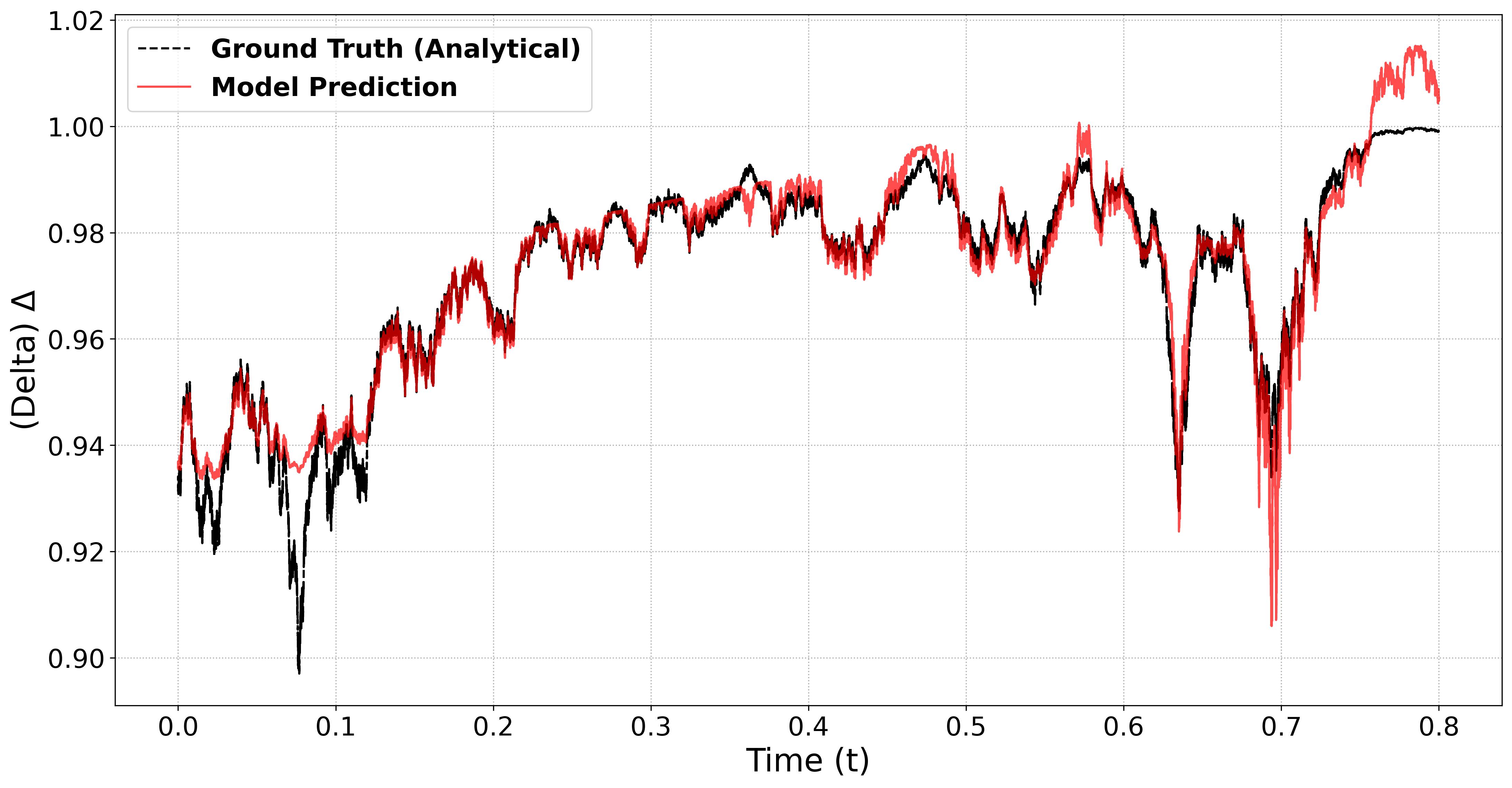}
\subcaption[third caption.]{\small Model vs analytical derivatives}\label{fig:1c}
\end{minipage}
\centering
\begin{minipage}{0.55\textwidth}
\includegraphics[width=0.8\textwidth]{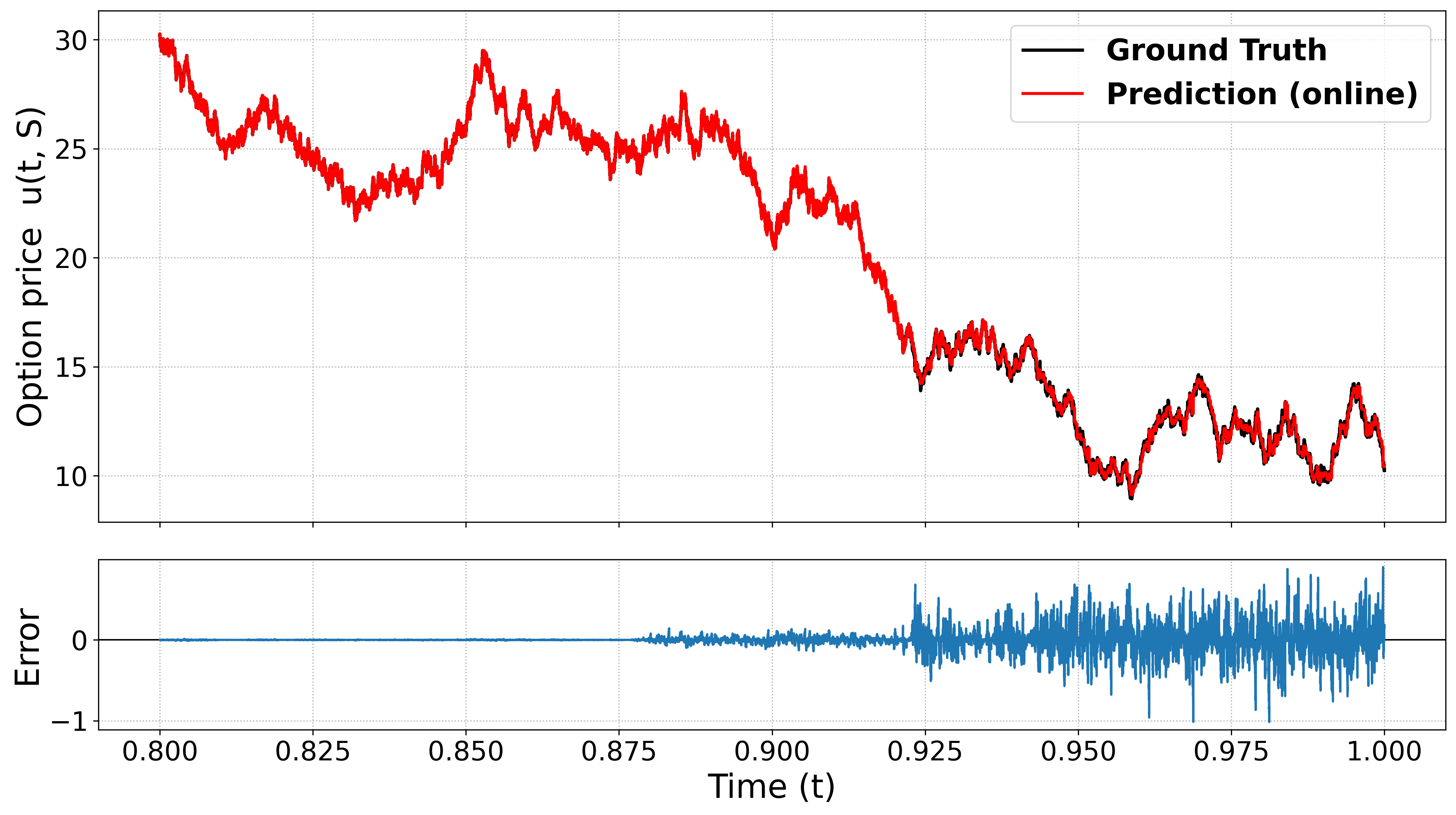}
\subcaption[second caption.]{Generation without retraining}\label{fig:1b}
\end{minipage}%
\begin{minipage}{0.55\textwidth}
\includegraphics[width=0.8\textwidth]{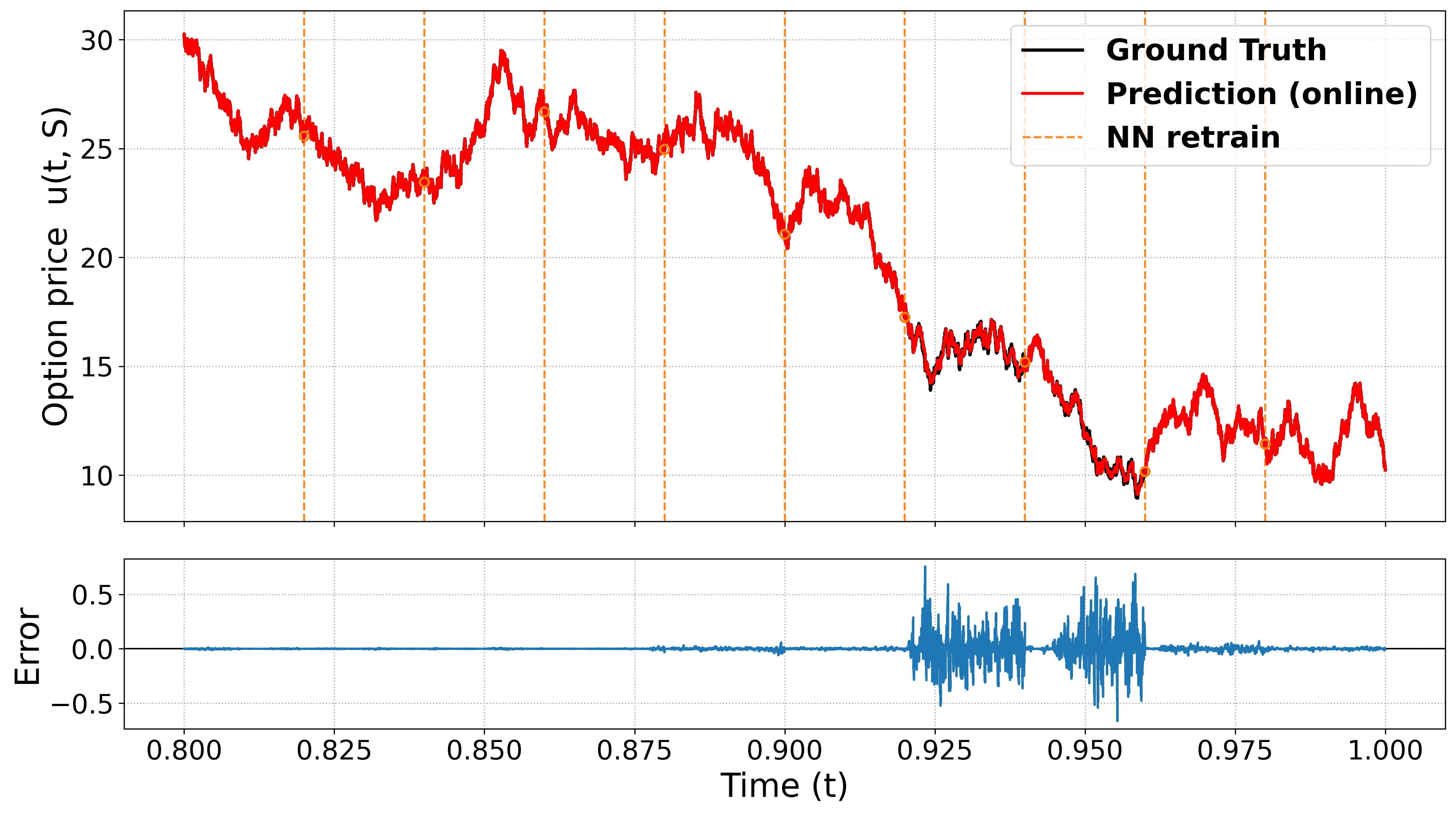}
\subcaption[third caption.]{Generation  with retraining}\label{fig:1c}
\end{minipage}
\centering
\begin{minipage}{0.33\textwidth}
\includegraphics[width=0.99\textwidth]{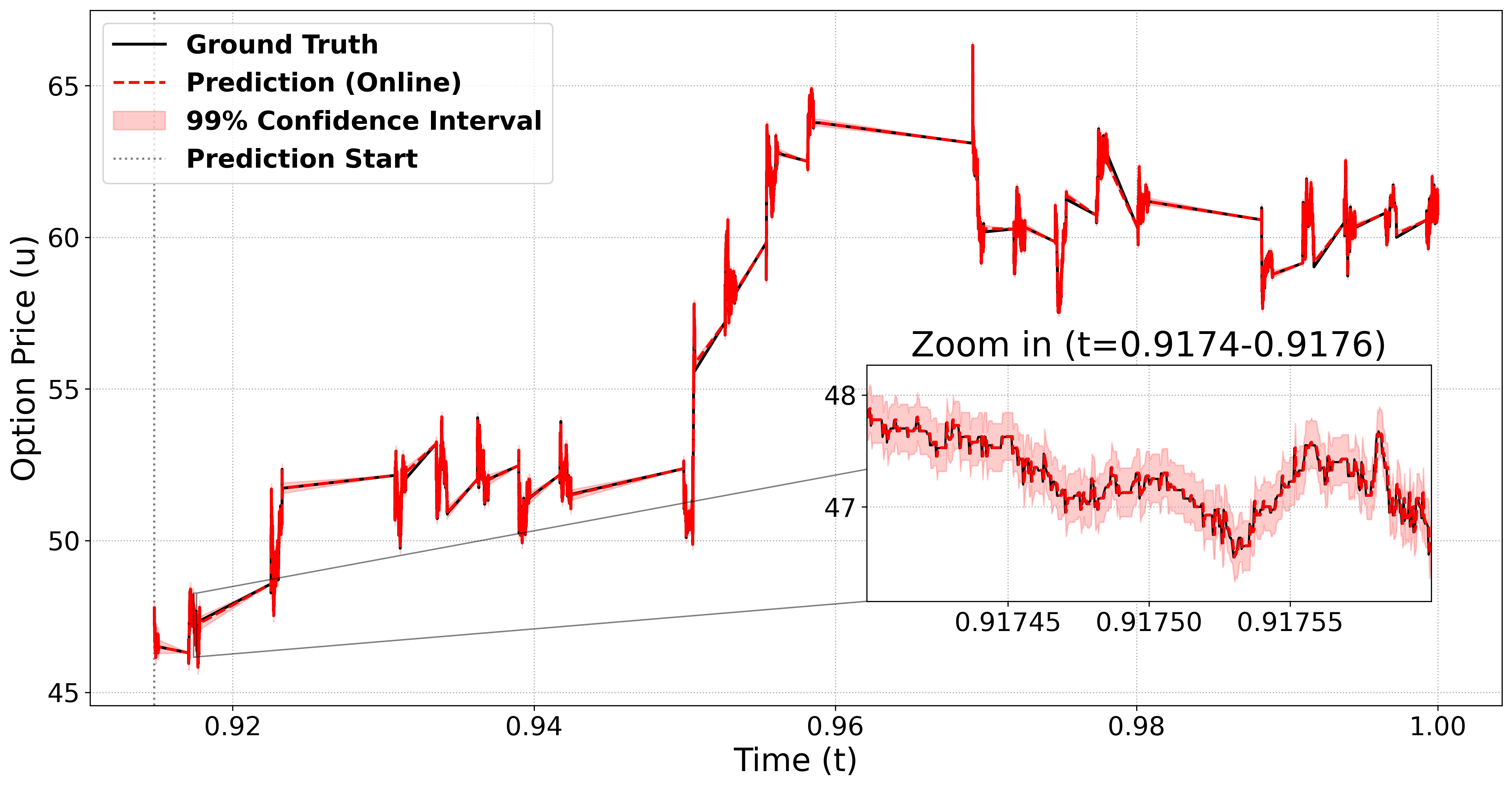}
\subcaption[second caption.]{Online prediction for Apple}\label{fig:1b}
\end{minipage}%
\begin{minipage}{0.33\textwidth}
\includegraphics[width=1.0\textwidth]{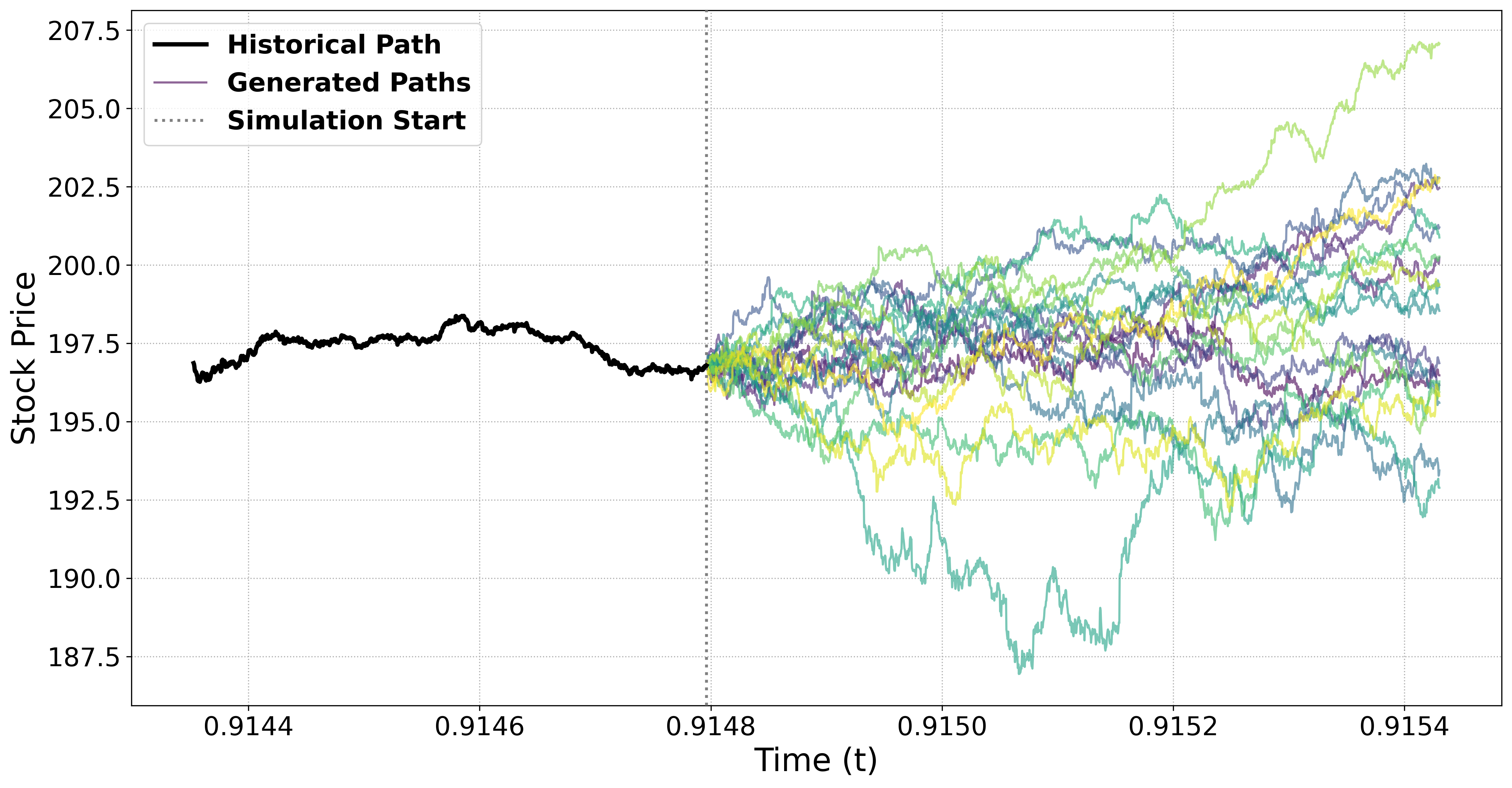}
\subcaption[third caption.]{Generation for Apple stock paths}\label{fig:1c}
\end{minipage}
\begin{minipage}{0.33\textwidth}
\includegraphics[width=1.0\textwidth]{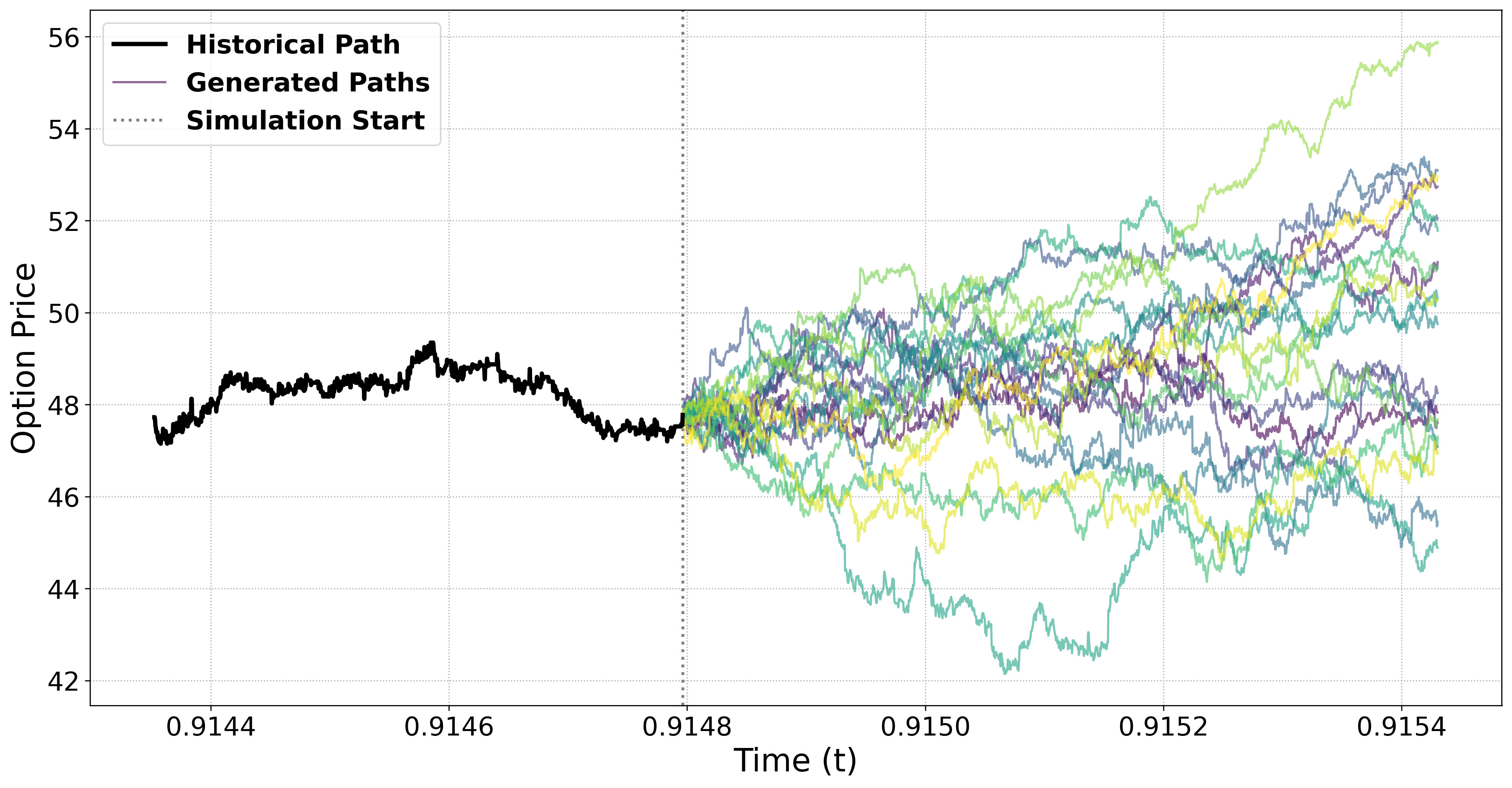}
\subcaption[third caption.]{Generation for Apple option paths}\label{fig:1c}
\end{minipage}

\caption{Numerical examples for Black-Scholes and Apple stock} \label{fig:1}
\end{figure}

\section{Conclusion}
We propose a stochastic SINDy algorithm for BSDE discovery (i.e., Feynman–Kac representation) under the risk-neutral probability measure, without requiring the ergodicity assumption. To the best of our knowledge, this is the first stochastic SINDy framework developed for backward stochastic differential equations, and the first algorithm to employ generalized stochastic SINDy to uncover probabilistic laws directly from a pair of financial time series—illustrated here with stock and option data. Beyond finance, this new algorithm points to a promising direction for the broader discovery of Feynman–Kac–type probabilistic laws in engineering and physics.

\section*{Acknowledgement}
Qi Feng is partially supported by the National Science Foundation under grant \#DMS-2420029. Denny Serdarevic is partially supported by the National Science Foundation under grant \#DMS-2420029 thorough the REU project at FSU. Guang Lin is partially 
supported by the National Science Foundation (NSF) (DMS-2533878, DMS-2053746, DMS-2134209, ECCS-2328241, CBET-2347401, and OAC-2311848), and DOE Office of Science Advanced Scientific Computing Research program DE-SC0023161, and DOE–Fusion Energy Science, under grant number: DE-SC0024583.
\bibliographystyle{plain}
\bibliography{refs, feng}

\end{document}